%% file: main.tex
\documentclass[10pt,conference]{IEEEtran}
\usepackage{cite}
\usepackage{graphicx}
\usepackage{xcolor}
\usepackage{booktabs}
\usepackage{makecell}
\usepackage{graphicx}
\usepackage{array}
\usepackage{colortbl}
\usepackage{amsmath,amssymb}
\usepackage{tikz}
\usepackage{tikz-cd}
\usepackage{multicol}
\usepackage[most]{tcolorbox}
\usepackage{array}
\usepackage[table]{xcolor}
\usepackage{colortbl}
\usepackage{hyperref}
\usepackage{pgfplots}
\usepackage{epigraph}
\usepackage{dirtytalk}
\pgfplotsset{compat=1.18}
\usepackage{xurl}

\usepackage[resetlabels,labeled]{multibib}
\newcites{EP}{Evaluated Papers}

\tcbset{
  mybox/.style={
    enhanced,
    colback=white,
    colframe=black,
    coltitle=black,
    colbacktitle=green!7,
    fonttitle=\bfseries,
    halign=flush left,
    boxrule=0.5pt,
    arc=2pt,
    left=6pt,
    right=6pt,
    top=6pt,
    bottom=6pt,
  }
}

\newcommand{\ignore}[1]{}

\begin{document}

\title{What Survives the Next Model? Benchmarking LLM-Based Techniques Against Single-Prompts}

\author{
\IEEEauthorblockN{Nahian Salsabil}
\IEEEauthorblockA{University of Virginia\\
abj6hv@virginia.edu}
\and
\IEEEauthorblockN{Joy Saha}
\IEEEauthorblockA{University of Virginia\\
daa7mv@virginia.edu}
\and
\IEEEauthorblockN{Simantika Bhattacharjee Dristi}
\IEEEauthorblockA{University of Virginia\\
nwc8gr@virginia.edu}
\and
\IEEEauthorblockN{Nicholas Phair}
\IEEEauthorblockA{University of Virginia\\
np4ay@virginia.edu}
\and
\IEEEauthorblockN{Nusrat Jahan Mozumder}
\IEEEauthorblockA{University of Virginia\\
nm8tm@virginia.edu}
\and
\IEEEauthorblockN{Matthew B. Dwyer}
\IEEEauthorblockA{University of Virginia\\
matthewbdwyer@virginia.edu}
\and
\IEEEauthorblockN{Sebastian Elbaum}
\IEEEauthorblockA{University of Virginia\\
selbaum@virginia.edu}
}


\maketitle

\begin{abstract}
The software engineering  research community has enthusiastically embraced the integration of Large Language Models (LLMs) into complex techniques to solve a wide variety of tasks. However, the extent to which this investment is strategic remains unclear, as the native capabilities of successive frontier model generations can rapidly render existing techniques obsolete. To assess this research investment, we analyze 35 LLM-based technique papers from ICSE 2026. We evaluate whether their complex tools can be outperformed by the simplest possible alternative: a single, automatically generated prompt executed on a newer generation model, without any iterative refinement. We find that for between 37\% and 63\% papers, a newer model with a single prompt natively outperforms the heavily engineered tooling proposed just a year prior. We identify that constructive techniques like code generation or repair are more amenable to substitution by a single-prompt. We also identify a surviving set of papers relying on strategies that provide additional insights to the model where newer LLMs will amplify the proposed technique. Our findings raise  questions about the cost-benefit proposition of techniques designed as workarounds to temporary model deficits and the need to focus on enduring challenges that scale synergistically with future model generations. Our source codes and results are made publicly available at \url{https://anonymous.4open.science/r/ICSE2027-What_Survives_The_Next_Model}.
\end{abstract}

\IEEEpeerreviewmaketitle

\nociteEP*

\input{intro}

\input{background}

\input{methodology}

\input{results}

\input{conclusion}

\clearpage

\bibliographystyle{IEEEtranS}
\bibliography{references}

\bibliographystyleEP{IEEEtranS}
\bibliographyEP{analyzedPapers}

\end{document}

%% file: intro.tex
\section{Introduction}
\begin{figure}
\includegraphics[width=1\columnwidth]{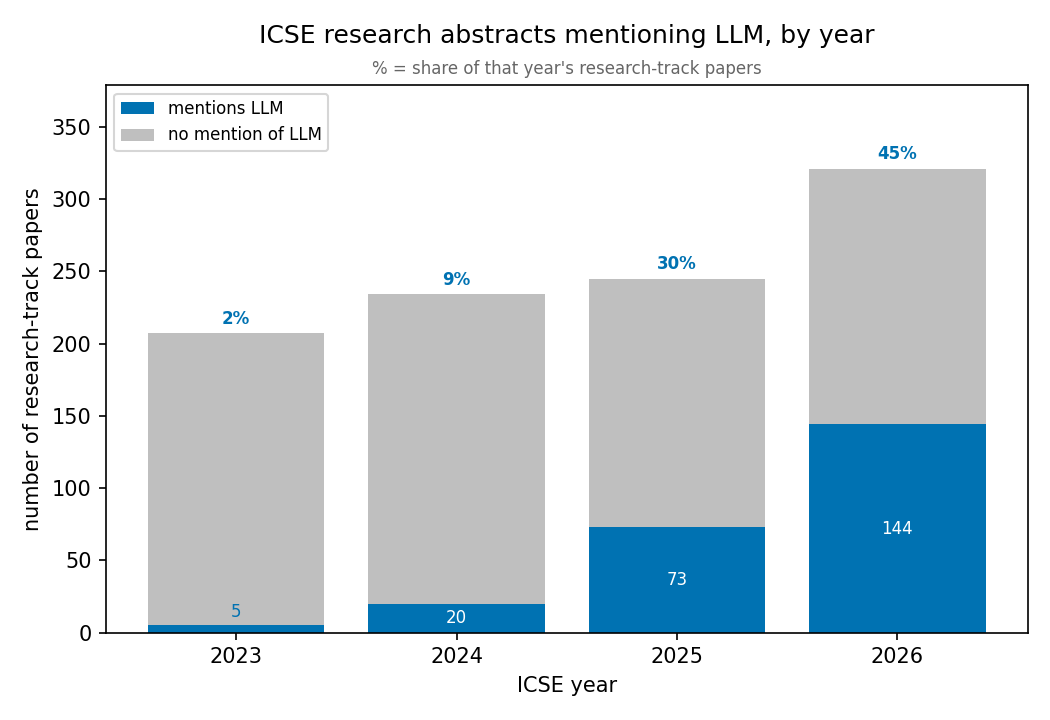}
\vspace{-0.2in}
\caption{ICSE Research Track papers with ``LLM'' in the abstract, 2023--2026. Each bar is the number of Research Track papers that year; the blue-shaded segment shows those whose abstract mentions ``LLM''.}
\label{fig:LLMinsabstracts} \vspace{-0.25in}
\end{figure}

We are seeing amazing growth in the ability of LLMs to solve  complex problems. The performance of frontier LLMs on programming challenges has been doubling year over year \cite{jain2025livecodebench,llmeval2024swebench}.
Beyond these raw performance gains, their applicability has also broadened to encompass the entire software development lifecycle \cite{2024systematicsurvey}. Industry has jumped in to adopt these techniques, with some reports indicating that 90\% of developers are using AI-based tools \cite{dora2025stateofai} and investment in ``tokens'' to use them has been exploding \cite{gartner2026tokens}.  The software engineering community has mirrored this enthusiasm. As shown in Figure~\ref{fig:LLMinsabstracts}, the fraction of ICSE Research Track papers mentioning ``LLM'' in their abstract grew from 2\% (5 of 207) in 2023 to 45\% (144 of 321) in 2026.

This enormous growth and sudden shift in research focus has similarities with a historical pattern in artificial intelligence research described by Richard Sutton in his essay, The Bitter Lesson \cite{bitterLesson}. As Sutton observed, \textit{``…[AI] researchers always tried to make systems that worked the way the researchers thought their own minds worked -they tried to put that knowledge in their systems- but it proved ultimately counterproductive, and a colossal waste of researcher's time, when, through Moore's law, massive computation became available and a means was found to put it to good use.’’} This insight warns that specialized human engineering designed to patch the immediate deficits of AI systems risks
being outpaced and rendered obsolete by scaling.

Today, the software engineering research community is at risk of experiencing a  manifestation of this lesson. While the scaling mechanism driving the current software engineering shift is no longer just raw hardware compute but the compounding capabilities of frontier LLMs, the underlying dynamic remains remarkably similar. As a stark illustration, consider the publication of AutoVerus \cite{yang2025autoverus} in October 2025, which established a new state-of-the-art in LLM-based verification of Rust code by incorporating a highly sophisticated collection of bespoke agents that codified human expertise, verification strategies, and iterative feedback loops. Yet, just two months later in December 2025, a new challenging benchmark for Rust verification was released \cite{yang2025verusage}. While AutoVerus successfully resolved 20\% of the problems, a newer frontier model (Sonnet 4.5), using a single, straightforward prompt, natively solved 81\% of the benchmark. A single generation of model improvement effectively deprecated the  engineering complexity of the specialized technique, while simultaneously yielding a four-fold performance improvement.

A single instance does not constitute proof across all software engineering domains, but it serves as a cautionary tale of a potential broader trend. We hypothesize that a substantial volume of current software engineering research may face a similar fate by allocating significant effort to two activities with a very short half-life: (1) identifying the quirky, localized limitations of a specific generation of LLMs and (2) engineering complex, bespoke workarounds to bypass them. Such short-term patches are highly likely to be rendered obsolete by the native  capabilities of future model generations, making them temporary fixes rather than enduring scientific contributions.
 
To test this hypothesis and quantify the extent of this pattern across the broader software engineering literature, we analyze the techniques of 35 LLM-based software engineering papers published in the most recent ICSE proceedings from 2026. We reproduce their key results and evaluate their core techniques against a bare baseline: a single, few-shot prompt submitted to a newer generation model than the one utilized in the original study. By design, our baseline represents the absolute minimum possible effort: the prompt is automatically generated directly from the paper’s text, involves no iterative refinement, requires no execution loops, and lacks access to external tools or infrastructure.

Our findings revealed a split in the longevity and value proposition of current LLM-based SE research. First, we observe a high ephemerality among complex bespoke techniques, finding that between 37\% and 63\% of the heavily engineered software engineering techniques using LLMs published less than six months ago can be effectively replaced by our single-prompt baseline. Constructive tasks, such as automated code generation or program repair, are the most susceptible to being natively absorbed by next-generation frontier models. For most of these techniques, the half-life is less than a single calendar year. We also discover that a single-prompt is insufficient with tasks requiring semantic understanding, complex feedback validation loops, or project-wide knowledge grounding. However, even for some of these domains, we project that simple additions, such as wrapping our single-prompts in a basic execution loop or granting them tool access, could bridge much of the remaining performance gap. 

On a positive note, our study highlights a resilient set of papers where model advancements do not replace the technique but rather amplify it. These papers provide a path forward for the community; they are characterized not by temporary workarounds, but by domain-specific processing, symbolic integration, and inputs structured specifically to provide richer context. 
Ultimately, this study provides the arguments to guardrail our research efforts from bridging temporary, fast-evaporating capability gaps in current LLMs. Instead, we should focus on enduring challenges that naturally scale with model capabilities rather than being erased by them.

%% file: background.tex
\section{Background}
\label{background}




Large language models (LLMs) are increasingly being used across many software engineering domains. This trend has been documented for several years. As early as 2023, Fan et al. published a survey \cite{ieeesurveyopenproblems} identifying potential LLM applications in coding, testing, requirements engineering, repair, refactoring, and documentation. More recent surveys show the realization of that potential. 
Hou et al. \cite{2024systematicsurvey} analyzed 395 SE papers from venues such as ICSE, FSE, and ASE, similarly documenting broad LLM adoption, for example, for code generation \cite{llmascodegen}, judging \cite{llmasjudge}, and agents \cite{llmasagent} that plan, critique, select, vote, or coordinate subtasks, and assessing optimization strategies, datasets, and evaluation mechanisms. Following these patterns, our study of 35 ICSE papers also cuts across tasks and strategies.


LLM usage across such a diverse set of tasks has driven the evolution of how we interact with LLMs in at least two ways. First, through the elevation of prompting to the level of a pseudo programming language for LLMs \cite{pseudoprog, promptprog}.
Multiple prompting surveys~\cite{llmprompteng, promptengsurvey, promptreport} catalog evolving prompting strategies (i.e., zero to few-shot, chain of thought to trees of thought) and offer templates that capture best practices \cite{template}.
Beyond prompting, the ability to leverage multiple LLMs with greater levels of autonomy has led to the emergence of agents. A recent survey \cite{2025comprehensive} 
shows that LLM-based SE techniques increasingly use agentic systems with planning, reasoning, memory, and tool augmentation. Against this backdrop, our study deliberately examines the simplest replacement strategy for software engineering techniques: a single-prompt to a single newer LLM, without access to any external tools and no multi-step interaction for iterative refinement.


Prior works \cite{llmeval2024swebench, llmevalbenchmark, llmevalrealworld} have also evaluated LLM capabilities for software engineering tasks, measuring how well a model performs on a particular benchmark or task in isolation. A recent systematic review \cite{llmeval}, for instance, characterizes how LLM capability is measured for standardized software engineering tasks by surveying 291 benchmarks. Their study explores these benchmarks, the SE tasks they cover, their construction methods, evaluation metrics, and remaining limitations in benchmark design. 
In this paper, we also perform a form of benchmarking. However, instead of purposefully defining different tasks or artifacts, we keep those fixed and introduce an alternative technique as treatment, substituting the original technique in a paper with a single-prompt to a newer LLM. As alluded to in previous work~\cite{muttakin2026state} and as we shall see in our study, performing this benchmarking in a systematic form across many papers is non-trivial. 

In this work we pose a different question:  will current LLM-based software engineering techniques become obsolete as newer models emerge?  This is not a question that has been asked, but it is critical, as the answer defines the operational half-life of the techniques we are developing.

%% file: methodology.tex
\newpage
\section{Research Questions}
We investigate the following research questions:

\begin{itemize}
    \item \textbf{RQ1:} Given a software engineering task, how effectively does single-prompt inference on state-of-the-art LLMs compete with a recent software engineering tool? 
    
    \item \textbf{RQ2:} What factors make a technique more amenable to single-call LLM substitution?
    
    \item \textbf{RQ3:} How much performance gain is achieved by embedding the original paper's methodology as reasoning guidance into the $P_w$ compared to $P_b$?


\end{itemize}

\section{Methodology}
\label{methodology}

\begin{figure}
    \centering
    \includegraphics[width=0.98\linewidth]{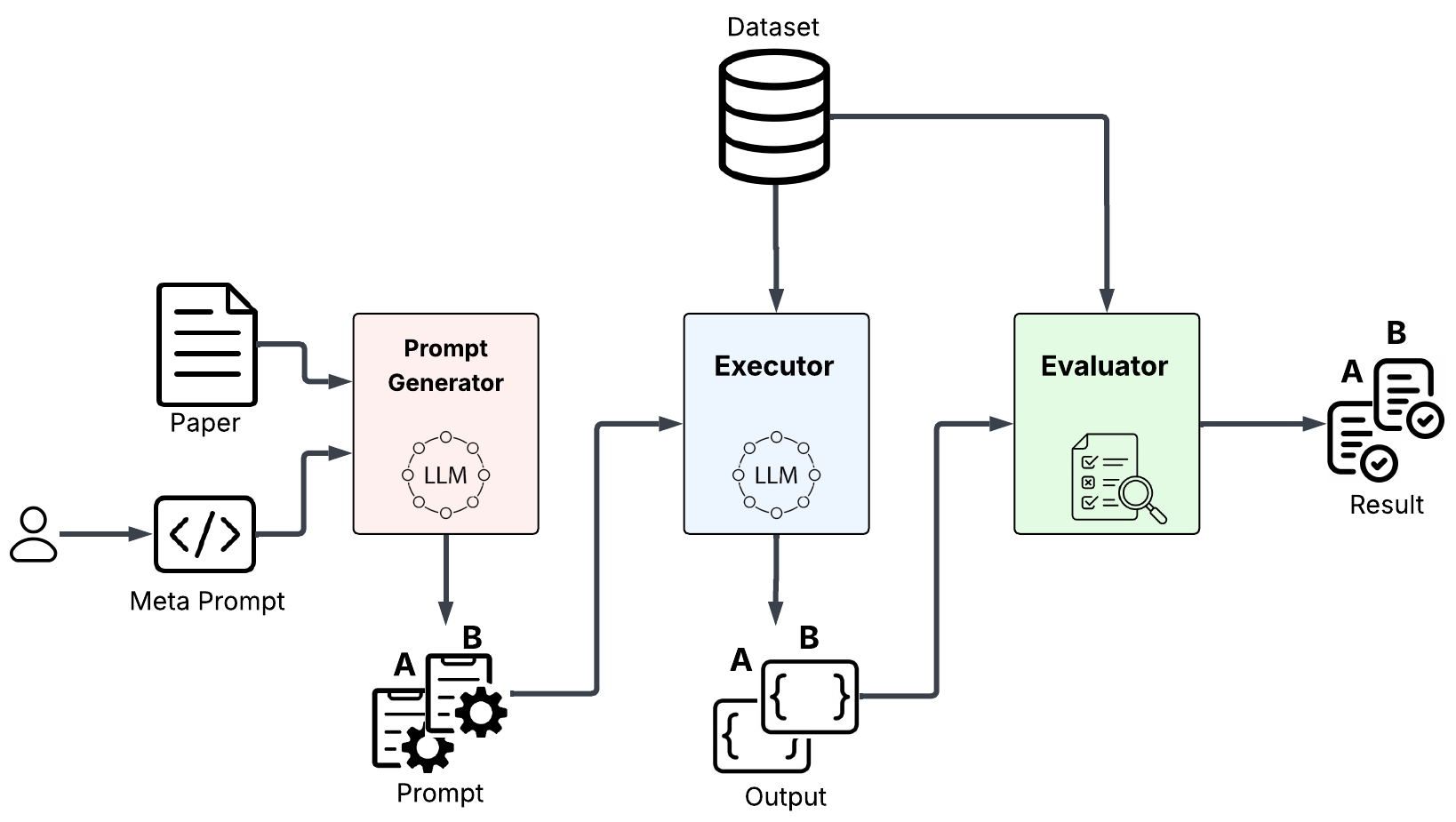}
    \caption{Methodology of single-prompt LLM approach}
    \label{fig:methodology}
    \vspace{-0.1in}
\end{figure}

This methodology, as shown in Figure \ref{fig:methodology} aims to provide evidence for a solid comparison between techniques with LLM components implemented in a subset of the ICSE 2026 papers and a single-prompt invoking a newer LLM  when applied to the same software engineering tasks. To carry out this comparison, we required the following: 

\begin{itemize}

\item A pool of candidate papers that implemented a technique using an LLM to solve a software engineering task and provided the necessary artifacts for evaluation of the prompt-based alternative.

\item A consistent mechanism to generate the single-prompts to solve each paper's task.

\item  An execution framework to run the generated prompts on the selected LLM with the same inputs and conditions established in the candidate papers.

\item An evaluation protocol to assess the prompt-driven LLM-generated outputs against the paper's tool over the same artifacts using the same process.

\end{itemize}


\subsection{Pool of Candidate Papers}

\begin{figure}
    \centering
    \includegraphics[width=0.96\linewidth]{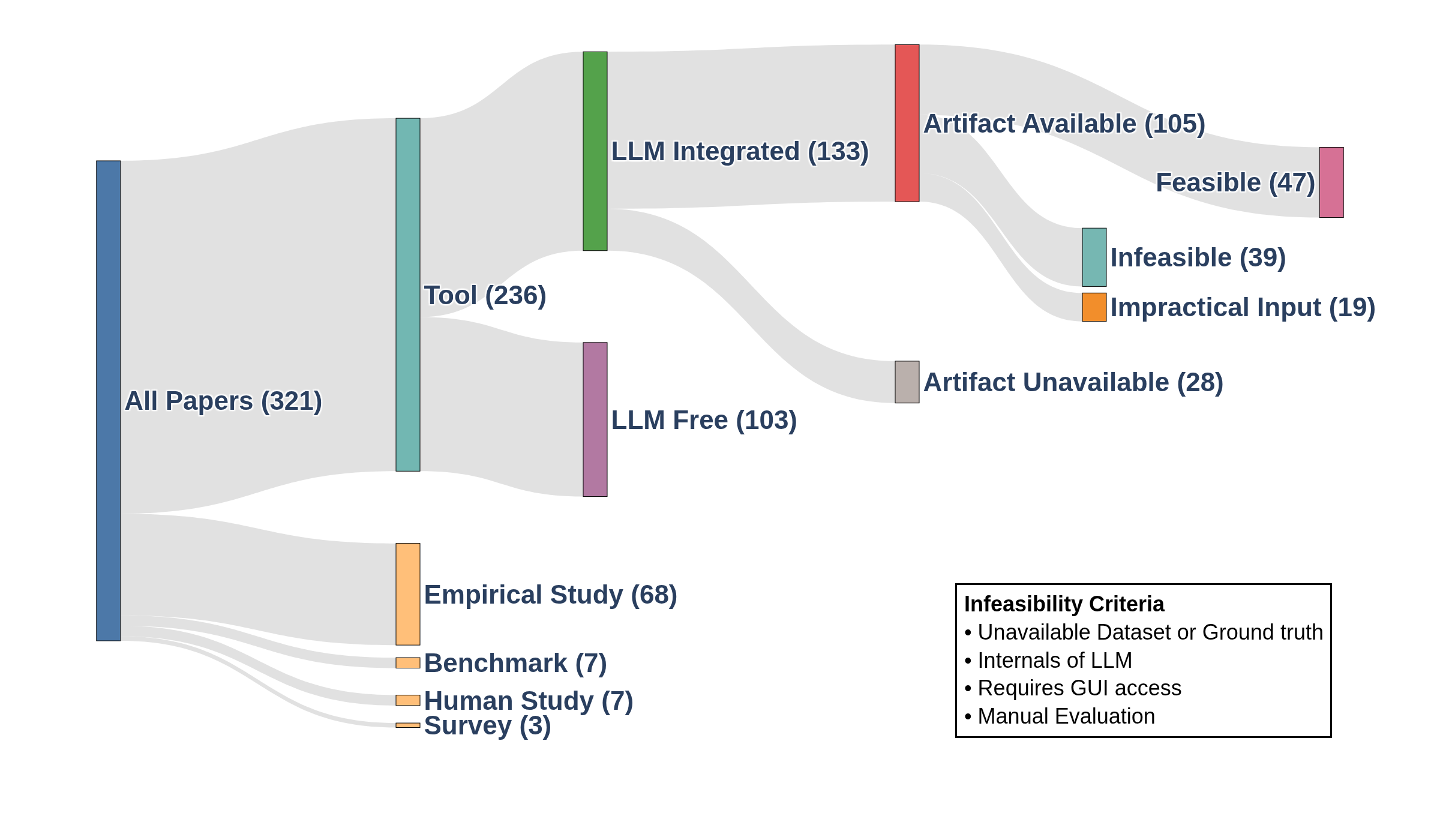}
    \caption{Paper selection process.}
    \label{fig:selection}
     \vspace{-0.1in}

\end{figure}

Figure~\ref{fig:selection} illustrates how we arrived at the pool of candidate papers, starting from the initial pool of 321 papers from the Research Track of the International Conference in Software Engineering 2026\cite{icse2026track},  the latest instance of the flagship conference in our field.

We first filtered 85 papers that did not introduce an automated technique to solve a task, this includes papers conducting empirical studies and surveys, developing benchmarks, or conducting human studies. Since the scope of our hypothesis is focused on papers with an LLM component, from the remaining 236 papers, we remove 103 papers that do not use an LLM. 
Out of the 133 remaining papers, only 105 
had publicly available artifacts to enable us to run our prompt-based approach. When inspecting their artifacts, we had to discard another 39 papers that did not provide the full dataset to perform the comparison, included an evaluation that required manual assessment, focused on the internals of the model instead of the software engineering task, or requiring specialized software (e.g., proprietary system, advanced simulator) to produce the input used as the data for evaluation. 
Next, we removed 19 papers that required extremely large inputs, as the cost of running such analysis is beyond our budget and LLMs are known to struggle in handling such large context \cite{lostinthemiddle}.  For example, \cite{icse2026_78} required full C source-code repositories as input to translate them into Rust. This requirement makes single-call substitution difficult because the dataset contains 25 repositories and roughly 71K C/C header files in total, with 18 repositories containing at least 100 such files. 
The processing cost for only the input would be on the order of roughly \$500, while including generated Rust output could raise the cost to several thousand dollars for a single pass over the full repository set.
After applying these exclusion criteria, we were left with a final pool of 47 papers. Of this final pool, we analyzed a random set of 35 papers in this study (cited with the prefix EP, and listed  under  `Evaluated Papers' in references). 



\subsection{Black-box and White-box Prompt Design and Generation}

To implement our single-prompt approach, for each paper, we needed prompts that instruct the LLM about the task it needs to perform, along with the corresponding inputs and outputs. We designed two types of prompts: Black-box ($P_b$) and White-box ($P_w$). These prompts share a common set of fields: \textit{Role, Task, Input, Output, Example input-output pairs, and Instructions}. The $P_b$ does not make any reference to the original paper's methodology, asking the LLM to infer the functionality to solve the task given the inputs and outputs provided. The $P_w$ extends the common structure with one additional field, \textit{Steps}, that embeds the paper's methodology as internal reasoning guidance, asking the LLM to realize the functionality following the paper's approach. This distinction allows us to examine whether providing the paper's methodology as guidance offers any additional benefit to the LLM's reasoning, a comparison that we later present in Section \ref{rq3}. 
The full structure of both prompts is shown in Table \ref{tab:prompt-structure}.


\input{tables/prompts_structure}

Given the two prompt designs, we set to automate their generation by designing a meta-prompt that, given a paper and input/output sample pairs, would invoke an LLM to generate paper-specific black-box and white-box prompts that instantiate the fields in Table \ref{tab:prompt-structure}. The design of this meta-prompt was iterative, refined using a set of 5 calibration papers until: the two prompts generated contained complete and identical content in every field except the Steps section, the $P_b$ did not include any methodological details from the paper,  and the required input and output format was correct. To address this, and throughout this process, we found that it was crucial to extract at least a single sample input/output pair from each paper's dataset, since the manuscripts themselves rarely detail these concrete execution artifacts. In general, we provided as few examples as possible but the number varied across papers.
For instance, \cite{icse2026_219} uses six datasets, and the expected output structure varies across them; therefore, we provided one example from each dataset. Similarly, in \cite{icse2026_186}, the output schema contains 15 fields, but no single dataset instance assigns meaningful values to all 15 fields. We therefore provided three examples whose combined outputs illustrate the full schema.

During this meta-prompt refinement we only evaluated the \textit{structure} of the generated prompts; we did not run experiments to assess the resulting prompts' performance on the target task and iterate on the meta-prompt. 
Once the meta-prompt satisfied the required criteria across the calibration papers, we adopted it as the single template used across all experiments. 
The template of the meta-prompt is shown in Figure \ref{fig:meta-prompt}.

\input{figures/meta-prompt}

\subsection{Prompt Executor}

 
This component invokes the LLM with the generated  prompts and the dataset from the candidate paper.

Before executing the prompt, we first manually checked that the papers' datasets satisfied the following criteria. First, 
they had to provide inputs and expected outputs against which we could check our results. If they did not, we removed the portions of the datasets that lack such ground truth. For example, \cite{icse2026_256} reports results on three datasets, but only one includes ground-truth vulnerability labels, thus we removed the other two datasets from our assessment.
Second, the inputs used by the paper's technique had to be readily available. If the inputs required, for example, any preprocessing or remote retrieval, we conducted those steps so that the generated prompts received the same inputs as the paper's technique. 
For example, \cite{icse2026_70} provides a set of Python programs but then it constructs its own evaluation dataset by removing certain function arguments from each file, stores the removed arguments as ground truth, and then asks their technique to recover them. We therefore replicated this step ourselves using Python's \texttt{ast} module to remove the corresponding arguments and produce the same incomplete-source inputs used by the paper's technique. 

Once the dataset met our criteria, our scripts made a single LLM call with the prompt and input. To control the execution costs of the study, we provided an LLM budget of \$15 per paper, which meant that not every dataset in every paper was fully analyzed. To accommodate this constraint fairly, we randomly sampled the inputs to analyze from each paper's dataset. Figure~\ref{fig:dataset-coverage} summarizes the percentage of each paper's evaluation dataset covered under this fixed budget. In the end, for 15 papers, that budget was enough to complete the analysis on the full dataset. For 16 papers, it covered 10\% to 70\% of the dataset. For 4 papers \cite{icse2026_6, icse2026_11, icse2026_32, icse2026_70}, we covered only 5\%--7\% and since the results were mostly negative, we did not pursue them further.  
We stored the model's raw response (for traceability) and the output in the requested structured json format. The formatted outputs (i.e., code, labels, test cases, constraints, prediction), were then evaluated against the original paper's expected outputs following its evaluation protocol.
\input{figures/dataset-percentage}
LLMs' capacities advance so rapidly that yesterday's frontier-leading model quickly becomes today's trailing model. For example, since the inception of this study, Anthropic has released several frontier models: Claude Opus 4.7 and Opus 4.8, Mythos, and Fable \cite{anthropic2026fable5}. For this study, we selected the trailing-edge Claude Sonnet 4.6 (released in February 2026). This decision was driven by three factors. First, Sonnet 4.6 represented the state of the frontier when we initiated this research; re-executing our entire approach on subsequent iterations would have required substantial renewed efforts and disrupted our established baseline. Second, Sonnet provides a non-trivial cost differential compared to leading-edge alternatives, as Anthropic's Opus models are approximately 1.67x more expensive than Sonnet for both input and output tokens, making cost a pragmatic consideration given the scale of our experiments across 35 papers. Third, this choice establishes a capability floor: if a trailing-edge model can successfully perform the target task, we can presume that future frontier-edge models will achieve at least the same performance.

\subsection{Evaluator}


This component aims to reproduce each paper's evaluation methodology to support a direct comparison between the paper-reported results and the results produced by our single-prompt approach. This phase consumed the most effort because each paper defines its own evaluation protocol that includes target metrics, ground truth, and execution environment.
To build each evaluator   we followed three steps. 

First, we identified the main research question (RQ). Papers typically pose multiple RQs. Since our goal is to compare against the tool's overall performance, we selected the RQ that most directly measures the end-to-end performance of the paper's technique. Then, for that  RQ, we identified the metrics used by the paper and adopted them in our evaluation to make our results directly comparable to the ones in the paper. 

Second, we prepared the ground truth needed to compute those metrics.   For 18 papers, the dataset provided explicit ground truth, such as labels, reference outputs, or affected program elements. In these cases, evaluation consisted of comparing the LLM-produced output directly against the provided ground truth using the paper's reported metrics. For the other 17 papers, the ground truth requires additional execution or checking the generated artifact. These cases required running tests, coverage tools, compilers, semantic equivalence checkers, or benchmark-specific evaluation harnesses before computing the final metric.

Third, we replicated the paper's evaluation setup so that the generated outputs were evaluated under the same conditions and metrics as the original technique. To give an idea of the complexity of this effort, we share a couple of examples. \cite{icse2026_14} generates tests to distinguish a plausible patch from an oracle patch, revealing behavioral discrepancies even when SWE-bench deems the patch correct. Checking whether a generated test differentiates the patches requires running it against both,  which implies we had to construct and launch a Docker container with an SWE-bench dataset, run the SWE-harness on both patches, and compare them. 
As another example, 
for \cite{icse2026_77}, which evaluates PostgreSQL-to-MySQL SQL dialect translation, we had to reproduce its execution-based setup by launching PostgreSQL and MySQL Docker containers, generating and loading identical 
database contents, running the source query on PostgreSQL and the translated query on MySQL, and comparing their result sets.

Overall, this evaluation process required a paper-specific evaluator for each of the 35 techniques, consuming several hours to understand, prepare, revise, and package. Each evaluator encoded the corresponding paper's metrics, ground truth format, and, when necessary, execution environment or external checking procedure. We make these evaluator scripts, along with the generated outputs and corresponding results, available in our replication repository.



\subsection{Artifact Quality Control and Paper Categorization}
 
We perform two additional cross-cutting processes to ensure artifact quality and enrich the analysis. First, we validate the correctness of the artifacts used in our experiments. Second, we label each paper on two technical factors that may affect whether a single-prompt can replace an existing technique: the type of software engineering task being performed and the key strategy used by the original technique to accomplish that task. 


To ensure the soundness of the produced artifacts, we performed an independent cross-reviewing process for each paper. We first defined a common checklist so that reviewers evaluated each artifact using the same criteria. Using this checklist, reviewers verified that the selected paper was appropriate for the study, the dataset was correctly identified, the evaluated research question matched the paper's main technique, and the metrics corresponded to that research question. They also verified that each prompt included at least one correct input-output example from the chosen dataset, the input-output formulation matched the paper's task, and the evaluator was correctly implemented to match each paper's evaluation setup, including the correct ground truth or the required execution environment. Each reviewer inspected multiple artifacts using this checklist. When a reviewer identified an issue, it was brought up for discussion and, as needed, corrected and reran before including it in the final analysis.

After completing the artifact checks, we categorized the papers
along the two factors introduced above: task type and strategy. To establish categories for these factors, rather than forcing papers into a predefined taxonomy, we utilized an incremental coding process. First, several reviewers independently characterized each paper's target domain and core technical aspects. A primary encoder then synthesized these characterizations alongside the   papers to explicitly code the stated software engineering domain and  technical contribution. We then iteratively grouped, refined, and consolidated these   codes to maximize distinctness and minimize overlap across our corpus. This process concluded once a consensus was reached among two reviewers, ensuring the resulting categories were sufficiently granular to capture technical key differences yet distinct enough to remain mutually exclusive.

This process resulted in the following  task labels according to software engineering tasks: code generation, bug finding, repair, impact analysis, requirement formalization, log analysis, test generation, and verification. Because some papers address more than one task or combine multiple technical strategies, we allow a paper to receive multiple task or strategy labels when appropriate. 
For strategy, the resulting labels require further explanation. \textit{Knowledge Grounding} refers to techniques that retrieve, construct, or inject additional project context, API relationships, dependency information, or static-analysis facts inside their approach. \textit{Feedback and Validation} refers to methodologies that use execution results, tests, compiler messages, traces, or other validation signals as feedback for other stages of the approach. \textit{Structured Reasoning} refers to techniques that decompose the task into explicit reasoning stages, agents, or hierarchical subtasks. \textit{Search and Selection} refers to approaches that generate multiple candidate solutions and rank, filter, or select among them. \textit{Input/Output Control} refers to frameworks whose main contribution is constraining, reducing, filtering, normalizing, or decoding the model's input or output schema. \textit{Domain-Specific Processing} refers to techniques that encode specialized rules, parsers, annotations, or procedures for a particular software engineering domain. \textit{Model Adaptation} refers to techniques that train, tune, or adapt a model or policy to improve performance on the target task.

%% file: tables/prompts_structure.tex
\begin{table*}[t]
\centering
\caption{White-Box and Black-box Prompt Structure}
\vspace{-0.1in}
\label{tab:prompt-structure}
\renewcommand{\arraystretch}{1.5}
\begin{tabular}{>{\raggedright\arraybackslash}p{0.08\linewidth} >{\raggedright\arraybackslash}p{0.20\linewidth} >{\raggedright\arraybackslash}p{0.62\linewidth}}
\arrayrulecolor{black}\hline\arrayrulecolor{black}
\textbf{Field} & \textbf{Description} & \textbf{Field Content} \\
\arrayrulecolor{black}\hline\arrayrulecolor{black}
Role & Task-specific LLM role. & You are an Expert software engineer specializing in API usage analysis and code ... \\
\arrayrulecolor{gray!25}\hline\arrayrulecolor{black}
Task & Brief task objective. & \makecell[tl]{You receive a code snippet with one or more missing API method arguments, indicated by a \\ `/* missing */` placeholder or similar comment... }\\
\arrayrulecolor{gray!25}\hline\arrayrulecolor{black}
Input & Input fields and definitions. & \makecell[tl]{\texttt{preceding\_code}: A string containing the full code snippet up...\\ \texttt{call\_line}: A string identifying the specific line or expression...} \\
\arrayrulecolor{gray!25}\hline\arrayrulecolor{black}
Output & Output schema and field definitions. & \makecell[tl]{A JSON object with a single key \texttt{arguments}, whose value is an array of objects, each containing:\\ \texttt{position}: integer\\ \texttt{value}: string} \\
\arrayrulecolor{gray!25}\hline\arrayrulecolor{black}
Example & Demonstrative input-output pairs. & \makecell[tl]{\texttt{inputs:}\\ \texttt{\ \ preceding\_code: "import os.path as path}\\ \texttt{\ \ \ \ \ \ \ \ \ \ \ \ \ \ \ \ \ \ \ PROJECT\_ROOT = path.realpath(path.dirname(\_\_file\_\_))}\\ \texttt{\ \ \ \ \ \ \ \ \ \ \ \ \ \ \ \ \ \ \ ...",}\\ \texttt{\ \ call\_line: "MEDIA\_ROOT = path.join(/* missing */)"}\\ \texttt{outputs:}\\ \texttt{\ \ arguments:}\\ \texttt{\ \ \ \ position: 0, value: "PROJECT\_ROOT"}\\ \texttt{\ \ \ \ position: 1, value: "media"}} \\
\arrayrulecolor{gray!25}\hline\arrayrulecolor{black}
Steps & White-box-only methodology steps. & \makecell[tl]{1. Identify the incomplete method call...\\ 2. Extract variable definitions and their types...\\ 3. Build an internal data-flow map...\\ 4. Reason about each missing argument position...\\ 5. Verify internal consistency...\\ 6. Compile and format the output...} \\
\arrayrulecolor{gray!25}\hline\arrayrulecolor{black}
Instructions & Fixed and task-specific constraints. & \makecell[tl]{- Solve the task using only the provided input.\\ - Do not use any external tools, APIs, web search...\\ - Provide a concrete value for every missing argument position...\\ - Output only valid JSON matching the schema above. Do not include explanation...\\ $\hdots$} \\
\arrayrulecolor{black}\hline\arrayrulecolor{black}
\end{tabular}
\vspace{-0.2in}
\end{table*}

%% file: figures/meta-prompt.tex
\tcbset{
  metabox/.style={
    enhanced,
    colback=gray!4,
    colframe=gray!60,
    coltitle=black,
    colbacktitle=gray!18,
    fonttitle=\bfseries\scriptsize,
    boxrule=0.45pt,
    arc=1mm,
    left=1mm,
    right=1mm,
    top=0.7mm,
    bottom=0.7mm,
    before skip=2pt,
    after skip=2pt
  }
}

\begin{figure}[t]
\centering
\begin{tcolorbox}[metabox, width=\columnwidth, title={Meta-Prompt for Single-prompt LLM Approach}]
\scriptsize

\textbf{Role:}
Expert AI Engineer specializing in Prompt Engineering and Technical Synthesis.

\vspace{0.25em}
\textbf{Task:}
Analyze a research paper and generate two self-contained replacement prompts,
Prompt Bb and Prompt Wb, that replace the paper's full multi-step technique with a single LLM call.

\vspace{0.25em}
\textbf{Rules for Both Prompts:}\\
-~Be fully self-contained... \\
-~The prompts must replace the paper's entire multi-step technique...\\
-~ ... \\
-~Embed the reference examples as a few-shot demonstration inside both prompts...

\vspace{0.25em}
\textbf{Steps vs. Instructions Boundary:}\\
-~Steps: The methodology...\\
-~Instructions: Behavioral constraints...\\
-~Do not duplicate content across Steps and Instructions.

\vspace{0.25em}
\textbf{Output Format:}\\
Provide both prompts as separate fenced Markdown code blocks.\\
Prompt Bb:\\
\texttt{\{See Table \ref{tab:prompt-structure} for structure\}}\\
Prompt Wb:\\
\texttt{\{See Table \ref{tab:prompt-structure} for structure\}}

\vspace{0.25em}
\textbf{Reference Examples:}
The following is a set of concrete input-output pairs representing exactly what the technique receives at the very start and produces at the very end...\\
\hspace*{0.8em}\texttt{"examples": [}\\
\hspace*{1.6em}\texttt{\{}\\
\hspace*{2.4em}\texttt{"inputs": \{}\\
\hspace*{3.2em}\texttt{"<input\_field\_1>": "<value>",}\\
\hspace*{3.2em}\texttt{"<input\_field\_2>": "<value>"}\\
\hspace*{2.4em}\texttt{\},}\\
\hspace*{2.4em}\texttt{"outputs": \{}\\
\hspace*{3.2em}\texttt{"<output\_field\_1>": "<value>"}\\
\hspace*{2.4em}\texttt{\}}\\
\hspace*{1.6em}\texttt{\}}\\
\hspace*{0.8em}\texttt{]}

\end{tcolorbox}
\caption{Meta-prompt template, together with the candidate paper, is sent to the LLM to generate $P_w$ and $P_b$.}
\label{fig:meta-prompt}
\vspace{-0.1in}
\end{figure}

%% file: figures/dataset-percentage.tex
\begin{figure}[t]
\centering
\resizebox{\columnwidth}{!}{%
\begin{tikzpicture}
\begin{axis}[
    ybar,
    width=0.98\textwidth,
    height=0.5\textwidth,
    ymin=0,
    ymax=105,
    ylabel={Dataset coverage (\%)},
    xlabel={Paper ID},
    symbolic x coords={
        20, 25, 38, 42, 54, 77, 97, 98, 103, 108, 186, 189, 219, 237, 256,
        73,18,10,82,254,232,19,102,104,52,86,88,14,66,303,154,6,70,11,32
    },
    xtick=data,
    x tick label style={rotate=90, anchor=east, font=\small},
    ytick={0,10,20,30,40,50,60,70,80,90,100},
    ymajorgrids=true,
    grid style={gray!25},
    bar width=5pt,
    enlarge x limits=0.015,
    tick style={black},
]
\addplot coordinates {
    (20,100)
    (25,100)
    (38,100)
    (42,100)
    (54,100)
    (77,100)
    (97,100)
    (98,100)
    (103,100)
    (108,100)
    (186,100)
    (189,100)
    (219,100)
    (237,100)
    (256,100)
    (73,75)
    (18,67)
    (10,56.5)
    (82,50)
    (254,50)
    (232,50)
    (19,37)
    (102,25)
    (104,25)
    (52,24)
    (86,17)
    (88,12.5)
    (14,10)
    (66,10)
    (303,10)
    (154,10)
    (6,6.25)
    (11,5)
    (32,5)
    (70,5)
    
};
\end{axis}
\end{tikzpicture}
}
    \vspace{-0.3in}
\caption{Percentage of each paper's evaluation dataset covered under the fixed LLM budget. Papers are sorted by coverage percentage.}
\label{fig:dataset-coverage}
    \vspace{-0.2in}
\end{figure}
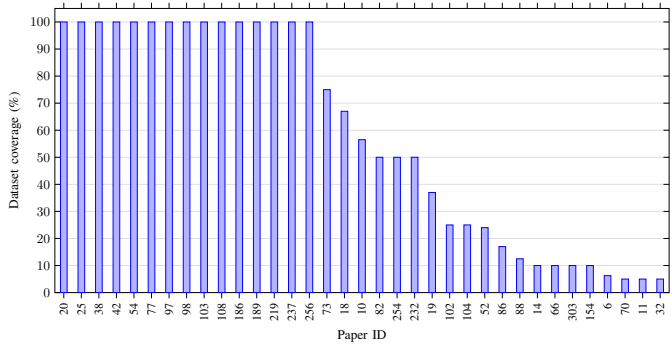

%% file: results.tex
\section{Results}
\label{results}

We now share the results for each research question.

\input{tables/results}

\subsection{RQ1: Single-Prompt LLM vs. Engineered SE Approaches}
\label{rq1}


Table~\ref{tab:results} summarizes the comparison between the original
techniques and the single-prompt replacements. 
The \textit{ID} column gives the paper identifier. The \textit{Paper's LLM} column reports the LLM used by the original technique. When a paper's technique is instantiated with multiple models, we pick the strongest model used for comparison. \textit{Dataset} and \textit{Metrics} identify the benchmark and paper-specific criteria used in our replication. \textit{Original Result} reports the paper's result on that dataset and metric.  $P_b$ and $P_w$ correspond to the prompt results. Finally, \textit{Diff(nl) Performance} compares the best single-prompt result against the original technique: positive ($+$) means the single-prompt fully outperformed the original technique, mixed ($+-$) when it outperformed the original only for a portion of the datasets or some but not all metrics, and negative ($-$) when it consistently underperformed the original. Bolded numbers indicate best performance in that row for a particular dataset portion and metric.

Overall, the results show that a single-prompt to a newer LLM is a surprisingly competitive substitute for many recent SE techniques engineered over older LLMs. Among the 35 studied papers, the single-prompt inference consistently outperformed the original technique in 13 cases and had mixed results in 9 additional cases.

\begin{tcolorbox}
\textbf{RQ1.}
Between 37\% and 63\% of recent SE techniques can be effectively replaced by an automatically generated single-prompt to a general-purpose LLM. \end{tcolorbox}

\begin{table}
\centering
\caption{Distribution of outcomes by task and strategy category.}
\label{tab:category-distribution}
\setlength{\tabcolsep}{2pt}
\renewcommand{\arraystretch}{1.2}
\begin{tabular}{|l|c|c|c||l|c|c|c|}
\hline
\textbf{Task Category} & \textbf{+} & \textbf{-} & \textbf{±} & \textbf{Strategy Category} & \textbf{+} & \textbf{-} & \textbf{±} \\
\hline
Bug Finding            & 1 & 7 & 3 & Domain-Specific Processing & 1 & 4 & 0 \\
Code Generation        & 6 & 2 & 1 & Feedback and Validation     & 3 & 3 & 3 \\
Impact Analysis        & 0 & 2 & 0 & Input/Output Control        & 2 & 2 & 0 \\
Log Analysis           & 1 & 1 & 1 & Knowledge Grounding         & 1 & 4 & 3 \\
Repair                 & 3 & 1 & 2 & Model Adaptation            & 1 & 1 & 1 \\
Req. Formalization     & 0 & 1 & 2 & Search and Selection        & 3 & 0 & 0 \\
Test Generation        & 1 & 2 & 1 & Structured Reasoning        & 5 & 1 & 3 \\
Verification           & 1 & 1 & 0 &                             &   &   &   \\
\hline
Total                  & 13 & 17 & 10 & Total                    & 16 & 15 & 10 \\ 
\hline
\end{tabular}
\vspace{-0.2in}
\end{table}

\subsection{RQ2: Amenability to single-prompt LLM substitution}
\label{rq2}

\begin{figure*}
    \centering
    \includegraphics[width=0.7\linewidth]{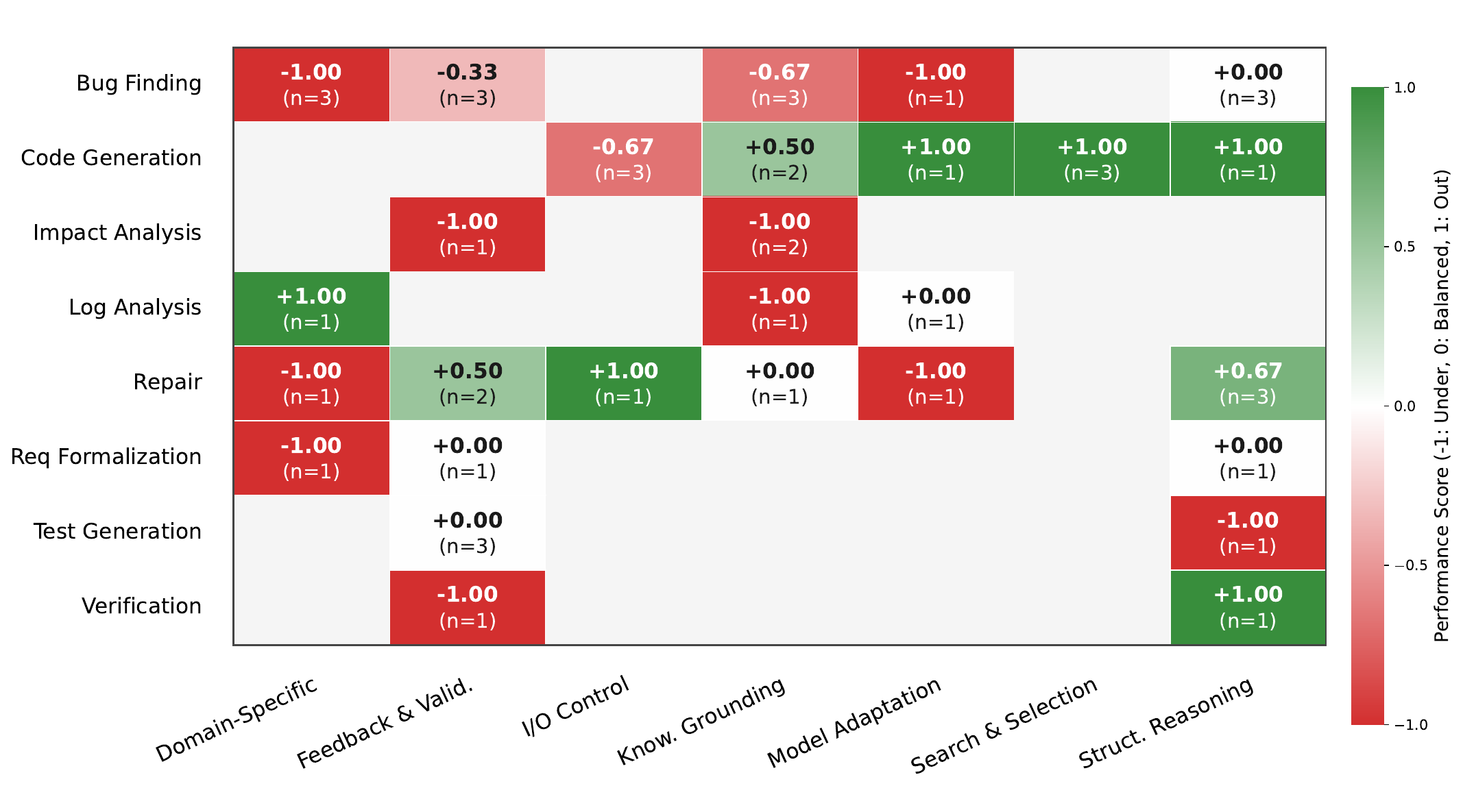}
        \vspace{-0.15in}
    \caption{Heatmap of tasks and strategies. Green cells show areas where the single-prompt outperforms the original technique, red indicates the opposite. $+1.0$ means that every single paper that used this strategy for this task showed that the single-prompt outperformed the technique in the paper. $0.0$ means  there was an equal balance of positive and negative.  $-1$ means every paper using this strategy for this task underperformed compared to the paper's technique. 
    }
    \label{fig:heatmap}
    \vspace{-0.2in}
\end{figure*}

Building on the paper categorizations introduced earlier, we explore how the target task type and the underlying technical strategies used by a given technique affect a single-prompt's viability as a replacement. We find that the effectiveness of the single-prompt varies substantially across task categories and strategies, but there are a few emerging patterns that we highlight 
through Table \ref{tab:category-distribution} and the heatmap in Figure \ref{fig:heatmap}.
In Table \ref{tab:category-distribution}, each row denotes either  a task or strategy category with columns denoting positive, negative and mixed results count per category. In Figure \ref{fig:heatmap}, each row represents a task category and each column a strategy. Green cells indicate the single-prompt outperforms the paper's technique, and red indicates the opposite. Cell's scores range from $-1.0$ (single-prompt underperforms in every paper using that strategy-task combination) to $+1.0$ (outperforms in every paper), where $0.0$ reflects a balance; n denotes the number of papers per cell, and empty cells lack papers for that combination. 

Papers performing constructive tasks such as code generation and program repair are more likely to be replaced by single-prompts.  As we can see in Table \ref{tab:category-distribution}, single-prompts for code generation outperform the techniques in six \cite{icse2026_10, icse2026_18, icse2026_19, icse2026_98, icse2026_108, icse2026_232} out of nine
papers, obtain mixed results for one \cite{icse2026_70}, and underperform in the other two \cite{icse2026_42, icse2026_77}.
For example, in \citeEP{icse2026_98}, the single-prompt achieved 78.85\% on pass@1, whereas the paper's original approach got 61.9\%. Code generation techniques that integrate strategies that constrain or guide the model were the easiest to outperform. For every instance of code generation using a `Search and Selection’ strategy, a single-prompt yielded consistently better results \cite{icse2026_18, icse2026_98, icse2026_108} (cell of code generation and `Search and Selection' in Figure \ref{fig:heatmap}). Overall, we find that newer LLMs excel at synthesizing functional code when they are anchored by simple constraints. 

Similarly, code repair techniques are also somewhat susceptible to quick and simple prompt-based replacements. As seen in Table \ref{tab:category-distribution}, for three of the six repair papers
a single-prompt outperforms the original technique \cite{icse2026_66, icse2026_73, icse2026_88}, and two results were mixed \cite{icse2026_86, icse2026_254}. Among them,
single-prompts were specifically well-suited to replace techniques that performed `Structured Reasoning' (two positive result \cite{icse2026_66, icse2026_73} and one mixed result \cite{icse2026_86} as seen in Figure \ref{fig:heatmap}, indicating that newer models may be able to capture those natively. For example, \cite{icse2026_66} uses structured reasoning through a three-agent repair workflow, where the agents translate the buggy program, generate repairs, and validate or revise the fix, yet the single-prompt replacement outperformed it. Surprisingly, single-prompts did not produce any negative result with one positive \cite{icse2026_66} and one mixed \cite{icse2026_86} on repair tasks even when the underlying techniques used `Feedback and Validation’ as a strategy. 

In contrast, single-prompts could not compete with tasks that require a deeper or specific semantic understanding. For example, as seen in the first row of Figure \ref{fig:heatmap} for Bug-finding tasks, single-prompts struggle
specially where the underlying techniques leverage 'Domain-Specific Processing’ \cite{icse2026_25, icse2026_186, icse2026_303}, `Knowledge Grounding’ \cite{icse2026_11, icse2026_154}, or `Feedback and Validation’ \cite{icse2026_237}.
This limitation extends to the Impact Analysis \cite{icse2026_32, icse2026_54} tasks as well,  as seen in the Table \ref{tab:category-distribution} having no positive or mixed results. The remaining tasks had fewer papers and did not show enough consistent patterns to report.

In Figure \ref{fig:heatmap}, looking at the strategies independent of the task, techniques relying on `Search and Selection' were the most replaceable, having all positive results, followed by `Structured Reasoning', 
As seen in Table \ref{tab:category-distribution}, all three papers \cite{icse2026_18, icse2026_98, icse2026_108} using `Search and Selection' outperformed the original technique. Of nine papers using structured reasoning, five outperformed \cite{icse2026_19, icse2026_20, icse2026_66, icse2026_73, icse2026_82}, three were mixed \cite{icse2026_86, icse2026_97, icse2026_219}, and one underperformed \cite{icse2026_103}. These results suggest that newer LLMs can often internalize such selection-based techniques and structured reasoning with a single-prompt. 

As seen in Figure \ref{fig:heatmap}, the hardest techniques to replace are those that relied on `Domain-Specific Processing' and `Knowledge Grounding' strategies. According to Table \ref{tab:category-distribution}, four of the five `Domain-Specific Processing' papers could not be outperformed by single-prompts. These techniques encode domain knowledge in task-specific procedures. For example, \cite{icse2026_38} is trained on dataset-specific translation guidelines by comparing generated formulas with ground-truth LTL, whereas our single-prompt approach receives no such dataset-level adaptation. 
This suggests that a single-prompt struggles when the original approach's advantage comes from specialized procedures learned or constructed for a particular domain. `Knowledge Grounding' showed a similar pattern,  with only one positive result  \cite{icse2026_10}, three mixed results \cite{icse2026_70, icse2026_254, icse2026_256}, and  four negative results \cite{icse2026_11, icse2026_32, icse2026_54, icse2026_154}, as the single-prompt approach could not derive such knowledge internally from the raw task input. These techniques generate intermediate knowledge or provide additional context, such as method dependencies \cite{icse2026_32}, static-analysis results \cite{icse2026_54, icse2026_154}, and API relationships \cite{icse2026_70},
and pass it to subsequent stages of the methodology. 
The remaining categories showed no consistent pattern.

We conjecture that two additional factors may contribute to these findings. First, the capability of the models used by the techniques in the papers. We expected techniques employing weaker models (see  ``Paper's LLM'' column in Table \ref{tab:results})
to be more susceptible to replacement, but the diversity of models, tasks, and strategies obscures any definitive pattern. Second, the evaluation datasets. Just a handful of popular benchmarks appear on multiple examined papers, so we state such trends with caution. We notice, for example, that the single-prompt outperformed every technique for the HumanEval \cite{humaneval} benchmark. This hints at that newer models may have utilized these benchmarks during training, however, the benchmarks’ age (e.g., HumanEval was first released in 2021) means older models likely consumed them as well. In other words, benchmark leakage could have happened in both cases, but we cannot assert how much it plays into the results. Future evaluations must keep this limitation in mind and adhere to the use of multiple contemporary benchmarks.

\begin{tcolorbox}
\textbf{RQ2.}
Single-prompt replacement is most effective for \textit{Code Generation} and \textit{Repair}, particularly with \textit{Search and Selection} and \textit{Structured Reasoning} strategies, but remains challenging for \textit{Bug Finding}, \textit{Program Analysis}, and \textit{Requirement Formalization}, especially when techniques rely on \textit{Knowledge Grounding} or \textit{Domain-Specific Processing}.
\end{tcolorbox}

\subsection{RQ3: The Value of White-Box Prompting}
\label{rq3}
\input{tables/prompt-comparison}

The white-box prompt extends the black-box prompt with guidance derived from the original
paper's technique. We expected this additional methodological guidance to make it more effective. However, as seen in Table \ref{tab:best-prompt-type} the results show a more balanced pattern. \textbf{$\mathbf{P_b}$ outperformed $\mathbf{P_w}$ in 14 of the 35 papers, while $\mathbf{P_w}$ outperformed $\mathbf{P_b}$ in 12 papers. The remaining 9 papers produced mixed results}, where one prompt was better on some datasets or metrics but not on all of them. Thus, the white-box guidance did not provide a consistent advantage over the simpler black-box prompt. 





A more detailed examination of the papers hints at why. Many of the techniques for which $P_b$ outperform $P_w$ depend on components or workflows that cannot be mimicked by the LLM. For example, \cite{icse2026_154} relies on static program-dependence and slicing analysis to trace and extract the parts of the smart contract code that influence signature checks before detecting replay vulnerabilities.  
Similarly, \cite{icse2026_256} uses static analysis to identify state dependencies, branch conditions, and function-call sequences; it then runs a fuzzer over those sequences, with runtime oracles built into the fuzzing process to detect when a vulnerability is actually triggered. These examples illustrate how having access to a description of the original procedure is insufficient for $P_w$ when the LLM cannot mimic or invoke the tools to generate the necessary data. 
We note again that by design we aimed to have the simplest alternative LLM-based implementation possible. As such, we do not allow the model to invoke tools or receive feedback, however, 
a natural next step is to explore agentic implementations, where the model can autonomously select and invoke external tools, the key capability that distinguishes agents from plain LLM prompting.

These additional instructions to guide $P_w$, however, do increase prompt complexity.
Across all papers, only for the prompt, $P_w$ requires an average    
of 80\% more tokens than $P_b$. We conjecture that such additional information, when the model cannot leverage it, acts as a distraction and is counterproductive, explaining the weaker $P_w$ results.
We did not find any pattern in the task types to explain the performance differences between $P_w$ and $P_b$. However,
at the strategy level, $P_b$ outperformed $P_w$ for `Knowledge Grounding' (5 out of 8, with 2 mixed), and slightly outperformed $P_w$  for `Domain-Specific Processing' (3 out of 5, with  none mixed),  and `Structured Reasoning' (3 out of 9, with 4  mixed). 
`Feedback and Validation' was the only strategy category where $P_w$ slightly outperformed $P_b$ (4 out of 9, with 2 mixed). 
These methodology descriptions often specify both the validation criteria and the feedback signal, such as test results, compiler errors, traces, or counterexamples. Although an LLM cannot execute these steps, we conjecture that  these validation details may help the model anticipate how its output will be judged, and the described feedback loop may force it to revise its answer internally before producing the final output.

We also compared the estimated dollar cost for running $P_b$ and $P_w$ that include the prompt, input, and output tokens. $P_b$ cost \$6.5 per paper on average, while $P_w$ cost \$7.5, with standard deviations of slightly over \$4 for both. Thus, $P_w$ was approximately 15\% more expensive than $P_b$. 


\begin{tcolorbox}
\textbf{RQ3.}
$P_b$ matches or outperforms $P_w$ in 40\% to 65\% of the studied SE tasks. In most cases LLMs cannot take advantage of the technical insights from the paper when provided through a single-prompt and without access to external tools.  
\end{tcolorbox}

\subsection{Threats to Validity}

\paragraph{External Validity}
Several factors may limit the generalizability of our findings. First, while our study examines a representative sample of papers, the selection from the ICSE proceedings does not capture the breath of software engineering, and  even less  of the practices in the broader community. Second, accessibility constraints to models driven by version decommissioning, high costs, or evolving geopolitical and regulatory restrictions may similarly affect generalizability. For instance, over the course of this study, model providers like Anthropic updated their model families, deprecate older models, and altered global access boundaries due to compliance and security protocols. We have mitigated these concerns by documenting our paper selection criteria and choosing contemporary, widely accessible model for our   evaluation.

\paragraph{Internal Validity}
Threats to internal validity relate to the inherent difficulties in performing a partial replication of highly diverse empirical studies across 35 distinct research groups. Given the scale of this task, although we aimed for standardization across prompt generation, execution setups, and evaluation pipelines, our replication can only guarantee a best-effort approximation. To minimize  errors, we implemented a cross-checking protocol involving multiple researchers that led to the refinement of many evaluations. Some of results depend on our categorization of techniques by task type and strategy; this classification scheme was adequate for our set of papers, but it may not be sufficient  more broadly. Budget constraints bounded the evaluation depth for a subset of the analyzed papers, which we mitigated by sampling. Finally, there is an unquantifiable threat regarding the evaluation benchmarks themselves; if the models were exposed to these datasets during training, their performance may be  inflated. We have explicitly documented the datasets to provide the necessary context for this threat.

\paragraph{Construct Validity} Threats to construct validity concern whether our broad characterization (outperforming, mixed, underperforming) captures the  performance  differences of the techniques studied. To enable a unified comparison over such a highly diverse set of techniques, datasets, and benchmarks, the constructs we chose are necessarily coarse. This characterization allows us to observe trends across heterogeneous studies, but they may lack nuance to uncover smaller patterns, or subtle qualitative dimensions unique to few techniques. To mitigate this limitation we commented on the possibility of such patterns when we found them, and we provide the core data in the paper and the rest in our repository to allow researchers to scrutinize these findings.

\paragraph{Conclusion Validity}
The inherent heterogeneity of the papers and their evaluation process, and the small sample size of papers, prevent us from conducting a statistical significance tests on our observations. We have tempered our conclusions accordingly, presenting our findings as indicative trends rather than definitive causal relationships.

%% file: tables/results.tex
\begin{table*}
\footnotesize
\centering
\caption{Comparison of paper-reported results against Single-Prompt LLM  
}  
\label{tab:results}
\resizebox{\textwidth}{!}{
\begin{tabular}{lllllllc}
\toprule
\thead{ID} & 
\thead{Paper's LLM} &
\thead{Dataset} &
\thead{Metrics} & 
\thead{Original Result} & 
\thead{$P_b$} & 
\thead{$P_w$} &
\thead{Diff(nl)\\Performance}\\
\midrule
\cite{icse2026_256} & Claude-3.5-Sonnet & EchoFuzz D2 & \#Vuln detected/TP/FP/FN
    & 103/103/\textbf{0}/8
    & 287/\textbf{111}/176/\textbf{0}
    & 297/\textbf{111}/186/\textbf{0}
    & $+-$ \\
\cmidrule{1-8}

\cite{icse2026_54}  & Claude-3.5-Sonnet & StatType-SO & F1
    & \textbf{90.8}
    & 42.9
    & 43.1
    & $-$ \\
\cmidrule{1-8}
\cite{icse2026_66}  & Claude-3.5-Sonnet & \makecell[l]{xCodeEval\\ (C, C\#, C++, Go,\\ Java, JS, Kotlin,\\ PHP, Python,\\ Ruby, Rust)} & Pass@5
    & \makecell[l]{90.42, 87.34,\\ 80.81, 90.46,\\ 85.99, 86.40,\\ 92.89, 99.23,\\ 90.66, 85.89,\\ 86.09}
    & \makecell[l]{\textbf{93.15, 93.15,}\\ \textbf{81.25,} 92.86,\\ 63.38, \textbf{100.00,}\\ \textbf{96.77, 100.00,}\\ \textbf{92.86,} 88.24,\\ \textbf{100.00}}
    & \makecell[l]{86.3, 90.41,\\ 81.25, \textbf{96.55,}\\ 66.20, \textbf{100.00},\\ 90.32, 89.92,\\ 89.47, \textbf{100.00,}\\ 90.00}
    & $\boldsymbol{+}$ \\
\cmidrule{1-8}
\cite{icse2026_25}  & CodeLlama-7B & \makecell[l]{Custom security code \\review dataset} & F1/SecureBLEU
    & \textbf{71.98}/\textbf{29.31}
    & 36.62/18.59
    & 32.97/16.59
    & $-$ \\
\cmidrule{1-8}
\cite{icse2026_73}  & DeepSeek-Coder-V2-Lite & HumanEval, MBPP & Accuracy/RSR 
    & \makecell[l]{94.5/76.3,\\ 80/39} 
    & \makecell[l]{\textbf{100}/\textbf{100} ,\\ \textbf{91}/\textbf{72.56}} 
    & \makecell[l]{98.17/92.11 ,\\ 90/69.51}
    & $\boldsymbol{+}$ \\
\cmidrule{1-8}
\cite{icse2026_97}  & DeepSeek-R1 & FoundRoot A-D & MRR
    & \makecell[l]{56.9, 61,\\ \textbf{85.8}, \textbf{93.1}}
    & \makecell[l]{61.33, 58, \\84.44, 92.52}
    & \makecell[l]{\textbf{64.85}, \textbf{61.79},\\ 83, 90.6}
    & $+-$ \\
\cmidrule{1-8}
\cite{icse2026_102} & DeepSeek-V3 & CodaMosa & \makecell[l]{(Line + Branch)\\ coverage}
    & \textbf{62}
    & 42.55
    & 46.17
    & $-$ \\
\cmidrule{1-8}
\cite{icse2026_154} & DeepSeek-V3 & DB2 (Ethereum) & F1
    & \textbf{88.46}
    & 40
    & 28
    & $-$ \\
\cmidrule{1-8}
\cite{icse2026_219} & DeepSeek-V3 & \makecell[l]{FSD, RAC, PURE,\\ BP, US, LMC} & \makecell[l]{N-F1/R-F1/,\\ Pass\_rate} 
    & \makecell[l]{\textbf{71.32}/\textbf{60.48}/\\ 97.78} 
    & \makecell[l]{48.53/49.99/\\ 96.34} 
    & \makecell[l]{49.65/50.71/\\ \textbf{98.38}}
    & $+-$ \\
\cmidrule{1-8}
\cite{icse2026_303} & DeepSeek-V3 & FSD & F1
    & \textbf{50.94}
    & 38.64
    & 35.71
    & $-$ \\
\cmidrule{1-8}
\cite{icse2026_88}  & DeepSeek-V3 & LFTBench & Pass@1
    & 45.9
    & 68
    & \textbf{72}
    & $\boldsymbol{+}$ \\
\cmidrule{1-8}
\cite{icse2026_38}  & DeepSeek-V3 & \makecell[l]{SynthNL, ConfoNL, \\LangNL, SpecNL} & Accuracy
    & \textbf{82}
    & 61.01
    & 58.18
    & $-$ \\
\cmidrule{1-8}

\cite{icse2026_104} & Gemini-2.5-Flash & \makecell[l]{FSM-AP, FSM-S, REG,\\ RobotExplain,\\ Ventilator, Deepstl-test} & Pass@10
    & \makecell[l]{\textbf{100}, \textbf{100}, \textbf{100},\\ \textbf{71.7}, \textbf{61.2}, \textbf{100}}
    & \makecell[l]{\textbf{100}, 66.7, \textbf{100},\\ 0, 20.7, 25}
    & \makecell[l]{\textbf{100}, \textbf{100}, \textbf{100},\\ 20, 27.6, \textbf{100}}
    & $+-$ \\
\cmidrule{1-8}
\cite{icse2026_86}  & Gemini-2.5-Flash & \makecell[l]{HumanEval, HumanEval+,\\ ClassEval, BigCodeBench} & Pass@1 
    & \makecell[l]{99.39, \textbf{98.17},\\ \textbf{82} , 85} 
    & \makecell[l]{\textbf{100}, 93.33,\\ 71, \textbf{87.72}} 
    & \makecell[l]{\textbf{100} , 93.33 ,\\ 74, 86.84}
    & $+-$ \\
\cmidrule{1-8}
\cite{icse2026_232} & Gemini-1.5-Flash & \makecell[l]{APPS, CodeContests,\\ xCodeEval} & Pass@1
    & \makecell[l]{65.33, 36.36,\\ 38.67}
    & \makecell[l]{\textbf{70.54}, 40.85,\\ \textbf{55.04}}
    & \makecell[l]{64.34, \textbf{43.66},\\ 48.06}
    & $\boldsymbol{+}$\\
\cmidrule{1-8}
\cite{icse2026_11}  & GLM4-2.0-Flash & \makecell[l]{BGL, Spirit, Thunderbird, \\HDFS, Hadoop2, Hadoop3, \\Spark2, Spark3} & F1
    & \textbf{97.5}
    & 18.5
    & 14.9
    & $-$ \\
\cmidrule{1-8}
\cite{icse2026_186} & GPT-4.1 & LV-Parser evaluation dataset & Mean accuracy 
    & \textbf{93.60} 
    & 82.99 
    & 86.77
    & $-$ \\
\cmidrule{1-8}
\cite{icse2026_98}  & GPT-4.1-mini & \makecell[l]{MultiPL-E Java} & Pass@1 
    & 61.90
    & \makecell[l]{\textbf{78.85}} 
    & \makecell[l]{77.88}
    & $\boldsymbol{+}$ \\
\cmidrule{1-8}
\cite{icse2026_32}  & GPT-4o & Alexandria & Hit@k/F1
    & \textbf{69}/\textbf{25}
    & 40.0/20.9
    & 40.0/17.3
    & $-$ \\
\cmidrule{1-8}
\cite{icse2026_103} & GPT-4o & AGORA+ & Precision/Recall
    & \textbf{85.1}/\textbf{83}
    & 53.7/41.6
    & 61.8/40.6
    & $-$ \\
\cmidrule{1-8}
\cite{icse2026_254} & GPT-4o & \makecell[l]{Defects4J V1.2,\\ Defects4J V2.0} & Pass test
    & \textbf{52.94}, 43.38
    & 43.33, 38.67
    & 46, \textbf{47.33}
    & $+-$ \\
\cmidrule{1-8}

\cite{icse2026_18}  & GPT-4o & \makecell[l]{HumanEval,\\ LiveCodeBench} & Pass@1
    & \makecell[l]{90.2, 50.2}
    & \makecell[l]{\textbf{99.1}, 52.51}
    & \makecell[l]{98.2, \textbf{52.8}}
    & $\boldsymbol{+}$ \\
\cmidrule{1-8}

\cite{icse2026_70}  & GPT-4o & \makecell[l]{Py150,\\ Netbeans} & Precision/Recall
    & \makecell[l]{\textbf{73.1}/68.1,\\ 72.11/69.46} 
    & \makecell[l]{66.12/65.47,\\ \textbf{72.7}/\textbf{72.57}} 
    & \makecell[l]{70/\textbf{69.06},\\ 71.5/71.38}
    & $+-$ \\
\cmidrule{1-8}
\cite{icse2026_82} & GPT-4o & PrimeVul & PC
    & 18.62
    & 13.37
    & \textbf{22.17}
    & $\boldsymbol{+}$\\
\cmidrule{1-8}
\cite{icse2026_237} & GPT-4o & \makecell[l]{PrOntoQA-OOD,\\ ProofWriter, FOLIO} & Average macro F1
    & \makecell[l]{\textbf{94.26}, \textbf{91.24},\\ \textbf{84.42}}
    & \makecell[l]{76.02, 65.79,\\ 56.82}
    & \makecell[l]{77.74, 84.28,\\ 66.66}
    & $-$ \\
\cmidrule{1-8}
\cite{icse2026_77}  & GPT-4o & \makecell[l]{TPC-DS,\\ SQLProcBench} & Translation accuracy
    & \textbf{97.98}
    & 60.61
    & 61.62
    & $-$ \\
\cmidrule{1-8}
\cite{icse2026_10}  & GPT-4o-mini & CoderEval & Pass@1
    & 36.52
    & \textbf{54.62}
    & 54.55
    & $\boldsymbol{+}$ \\
\cmidrule{1-8}
\cite{icse2026_108} & GPT-o3-mini & GuideSyn .mls benchmark & \%solved
    & 78
    & 81.25
    & \textbf{83.75}
    & $\boldsymbol{+}$ \\
\cmidrule{1-8}
\cite{icse2026_14}  & GPT-4o mini & SWE-bench & Differentiating rate 
    & \textbf{29.3}, 27.2
    & 15.56 ,\textbf{36.73} 
    & 11.11, 24.49
    & $+-$ \\
\cmidrule{1-8}
\cite{icse2026_52}  & Llama 3.3 70B & Defects4J & \makecell[l]{Line/Branch\\ coverage, Pass rate}
    & \makecell[l]{78.62/69.25,\\ 62.91}
    & \makecell[l]{89.3/80.3,\\ \textbf{98.9}}
    & \makecell[l]{\textbf{91.7}/\textbf{82.9},\\ 98.4}
    & $\boldsymbol{+}$ \\
\cmidrule{1-8}
\cite{icse2026_6}   & Qwen2.5-14B & HPC (Loghub-2k) & PA/PTA/RTA/GA
    & \textbf{99.3}/71.7/\textbf{82.6}/\textbf{93.4} 
    & 90.5/72.9/76.1/90
    & 90.5/\textbf{78.3}/78.3/90.5
    & $+-$ \\
\cmidrule{1-8}
\cite{icse2026_189} & Qwen2.5-32b-instruct & \makecell[l]{Custom GitHub \\derailment dataset} & F1@threshold 0.3
    & 90.10
    & 85.82
    & \textbf{90.91}
    & $\boldsymbol{+}$ \\
\cmidrule{1-8}
\cite{icse2026_20} & Qwen2.5-72B & CoCoClaNeL & MCC
    & 25.9
    & \textbf{65.72}
    & 65.41
    & $\boldsymbol{+}$\\
\cmidrule{1-8}
\cite{icse2026_19}  & Qwen2.5-Coder-7B-Instruct & \makecell[l]{HumanEval, MBPP,\\ LiveCodeBench} & Pass@1
    & \makecell[l]{90.9, 88.6,\\ 36.7}
    & \makecell[l]{\textbf{100}, \textbf{100},\\ 48.68}
    & \makecell[l]{\textbf{100}, 99.29,\\ \textbf{50.94}}
    & $\boldsymbol{+}$ \\
\cmidrule{1-8}
\cite{icse2026_42}  & \makecell[l]{StarCoder2-7B (v1),\\ CodeLlama-Python-7B (v2)} & Tfv1 Synthetic & Exact match@1
    & \makecell[l]{\textbf{83.53} (v1),\\ \textbf{60.86} (v2)}
    & \makecell[l]{71.4 (v1),\\ 36.5 (v2)}
    & \makecell[l]{70.9 (v1),\\ 36.3 (v2)}
    & $-$ \\
\bottomrule
\end{tabular}
}
\end{table*}

%% file: tables/prompt-comparison.tex
\begin{table}[t]
\centering
\caption{Best-performing prompt type across papers.} 
\label{tab:best-prompt-type}
\footnotesize
\setlength{\tabcolsep}{4pt}
\renewcommand{\arraystretch}{1.15}
\begin{tabular}{
    |>{\raggedright\arraybackslash}p{0.25\columnwidth}
    |>{\raggedright\arraybackslash}p{0.5\columnwidth}
    |>{\centering\arraybackslash}p{0.14\columnwidth}|
}
\hline
\textbf{Best Prompt Type} & \textbf{Paper IDs} & \textbf{Count} \\
\hline
$P_b$ &
\cite{icse2026_10}, \cite{icse2026_11}, \cite{icse2026_14}, \cite{icse2026_20}, \cite{icse2026_25}, \cite{icse2026_32}, \cite{icse2026_38}, \cite{icse2026_42}, \cite{icse2026_66}, \cite{icse2026_73}, \cite{icse2026_98}, \cite{icse2026_154}, \cite{icse2026_256}, \cite{icse2026_303}&
14 \\
\hline
$P_w$ &
\cite{icse2026_18}, \cite{icse2026_54}, \cite{icse2026_77}, \cite{icse2026_82}, \cite{icse2026_88}, \cite{icse2026_102}, \cite{icse2026_108}, \cite{icse2026_186}, \cite{icse2026_189}, \cite{icse2026_219}, \cite{icse2026_237}, \cite{icse2026_254} &
12 \\
\hline
Mixed &
\cite{icse2026_6}, \cite{icse2026_19}, \cite{icse2026_52}, \cite{icse2026_70}, \cite{icse2026_86}, \cite{icse2026_97}, \cite{icse2026_103}, \cite{icse2026_104}, \cite{icse2026_232} &
9 \\
\hline
\end{tabular}
\vspace{-0.2in}
\end{table}

%% file: conclusion.tex
\section{Conclusion}
\label{conclusion}

This paper is the first to systematically investigate the longevity and strategic value of software engineering techniques built around LLMs. Our study of 35 SE papers from ICSE 2026 makes it recent and thus relevant; our automated prompt-generation, execution, and evaluation framework ensures that our methodology is both systematic and reproducible. That said, we acknowledge that this work has limitations, particularly regarding fixed execution budgets, dataset sampling constraints, and categorization schemas, and that it represents just the first step in this line of inquiry.

Our findings reveal that between 37\% and 63\% of SE techniques published less than six months ago can be effectively replaced by the simplest possible instance of LLM invocation: a single, automatically generated prompt executed on a newer generation model, without any iterative refinement or access to other tools. These findings provide a warning to the software engineering research community about the ephemeral nature and short half-life of many of the heavily engineered workarounds our community is currently producing. 

Our findings also reveal that techniques aimed at constructive tasks like generating code or repairing code
are the easiest to be replaced as frontier models get more powerful. For tasks requiring deeper semantic understanding or relying on  feedback loops, existing techniques still fare better, but  some of the performance gains could be matched if the single prompts are simply placed within a basic loop, delegated to agents, or granted access to other tools, something that future work should explicitly investigate. On the other hand, papers involving domain-specific processing or project-wide knowledge grounding are the least likely to be replaced.


Overall, these insights point to the critical need for our community to be far more strategic in our research investments. We are currently relearning AI’s `Bitter Lesson’ within the software engineering domain: heavily engineered human heuristics designed to patch temporary model deficits are rapidly cleared away by computational scaling. Enduring software engineering research must shift from building transient scaffolds for LLMs towards establishing where symbolic approaches 
can add complementary value.
Our community needs to stop chasing the moving target of model capabilities and instead build techniques that scale alongside them.